\documentclass[aps,preprint]{revtex4-1}
\usepackage{graphicx}
\usepackage{amsmath}
\usepackage{makeidx}
\usepackage{amsfonts}
\usepackage{amssymb}

\begin{document}

\title{Cavity approach to variational quantum mechanics}

\author{A. Ramezanpour}
\email{abolfazl.ramezanpour@polito.it}
\affiliation{Physics Department and Center for Computational Sciences, Politecnico di Torino, Corso Duca degli Abruzzi 24, I-10129 Torino, Italy}

\date{\today}

\begin{abstract}
A local and distributive algorithm is proposed to find an optimal
trial wave function minimizing the Hamiltonian expectation in a
quantum system. To this end, the quantum state of the system is
connected to the Gibbs state of a classical system with the set of
couplings playing the role of variational parameters. The average
energy is written within the replica-symmetric approximation, and the optimal
parameters are obtained by a heuristic message-passing algorithm
based on the Bethe approximation. The performance of
this approximate algorithm depends on the structure and quality of the
trial wave functions, starting from a classical system of isolated
elements, i.e., mean-field approximation, and improving on that by
considering the higher-order many-body interactions. The method is
applied to some disordered quantum Ising models in transverse fields,
and the results are compared with the exact ones for small systems.
\end{abstract}


\maketitle

\section{Introduction}\label{S0}
The cavity method, relying on the Bethe approximation \cite{B-prs-1935}, was originally introduced as an alternative to 
the replica method to study equilibrium properties of disordered and effectively mean-field classical systems \cite{MPV-book-1987,MP-epjb-2001}. 
Later on it was also considered as a powerful message-passing algorithm to solve for the solutions in single instances of 
some computationally difficult problems \cite{MM-book-2009,MPZ-science-2002,BMZ-rsa-2005}. Recently, we applied the method
to a class of more challenging optimization problems, where the objective function itself is a computationally complex
function, e.g., the average of a minimum energy function in a stochastic optimization problem \cite{ABRZ-prl-2011,ABRZ-jstat-2011}.
In this study we are using these advancements to develop a message-passing algorithm for
variational quantum-mechanics problems.

Cavity-like approaches to quantum systems differ in nature and scope: 
The quantum belief propagation algorithms \cite{H-prb-2007,LP-aphys-2008,PB-pra-2008} are the quantum generalization of 
the classical belief propagation algorithm \cite{KFL-inform-2001} working with local cavity density matrices instead of the 
cavity marginals. The smaller the temperature is, of course, the larger the density matrices needed are as quantum 
correlations prevail in the system. On the other hand, the quantum cavity method 
\cite{CWW-prb-1992,LSS-prb-2008,KRSZ-prb-2008,STZ-prb-2008,IM-prl-2010} 
maps the quantum problem to a classical one using the Suzuki-Trotter transformation, and then exploits the cavity method to 
estimate the relevant average quantities. This can be regarded as a dynamical mean-field theory extended to take the spatial 
correlations into account. 

In this work we take another approach that merges the techniques we used in the stochastic optimization problems \cite{ABRZ-prl-2011,ABRZ-jstat-2011} with 
the variational principles of quantum mechanics, i.e., the fact that any trial wave function (density matrix) provides 
an upper bound for the ground-state energy (free energy). 
Alternatively, a lower bound for the free energy can be obtained by approximating the entropy with an overestimated entropy function 
\cite{PH-prl-2011}. 
In both cases, the problem of finding the optimal state can be recast as an optimization problem with an objective 
function that evaluates the (free) energy of a physical state. In the rest of this paper we shall focus on the simpler 
problem of finding an optimal wave function, i.e., at zero temperature.     

The strategy of finding the optimal wave function in a quantum system is analogous to that of finding the optimal 
configuration in a classical optimization problem. Both the problems are, in general, computationally hard, making efficient 
and accurate heuristic algorithms central to the study of these problems. The quantum problem is intractable already in one 
dimension \cite{AGIK-mphys-2009} or for fermionic systems due to the sign problem \cite{TW-prl-2005}. 
In addition, it is important for the efficiency of the variational method, to have a succinct representation of the trial wave functions 
that accurately describes the ground state of a quantum system \cite{H-prb-2006,VMC-aphys-2008}. 

In the following we shall map the quantum wave function to the Gibbs state of a classical system, considering the set of 
couplings as the variational parameters. The first step of the algorithm is to choose an appropriate classical system that 
best captures the quantum nature of the original system. The minimal set of couplings is
that of a classical system of isolated elements, and in the maximal set one would have the whole set of many-body interactions. 
The former is equivalent to the mean-field approximation
and the latter to an exact treatment of the problem. However, as we will see, already the two-body interactions give 
a reasonable estimate of the physical quantities in a disordered quantum Ising model. 
One may compare this with the Jastrow trial wave function, which is the product of pair functions \cite{J-pr-1955}.   

The second step of the algorithm is to write the objective function, which is the quantum average of the Hamiltonian, 
in terms of classical and local average quantities estimated within the replica-symmetric approximation 
\cite{MM-book-2009}. The quality of this approximation depends on the structure of the classical interaction graph; 
it is expected to work well on random and sparse graphs, that is in effectively mean-field systems. 
In general, however, this approximation spoils the upper bound property of the Hamiltonian expectation we started from.      
The third and last step of the algorithm is to find the optimal couplings and we do this by a heuristic message-passing 
algorithm based on the Bethe approximation.  

This paper is organized as follows. We start in section \ref{S1} with the variational quantum problem in its general 
form and write it as a classical optimization problem amenable to the cavity method.
In section \ref{S2} we write the cavity equations for a quantum spin model and compare the different levels of approximations with 
the exact results. Section \ref{S3} gives the concluding remarks.

\section{General arguments}\label{S1}
Given a Hamiltonian $H$ and a trial wave function  $|\psi(\underline{P})\rangle$, we have $\langle \psi(\underline{P}) | H |\psi(\underline{P})\rangle \ge E_g$
where $E_g$ is the ground-state energy of $H$ and $\underline{P}$ denotes a set of parameters characterizing the trial wave function.
To find the optimal parameters we define the following optimization problem:
\begin{eqnarray}
\mathcal{Z}=\sum_{\underline{P}} e^{-\beta_{opt} \langle \psi(\underline{P}) | H |\psi(\underline{P}) \rangle },
\end{eqnarray}
where eventually one is interested in the limit $\beta_{opt} \to \infty$. Note that $\beta_{opt}$ is just the inverse of a fictitious temperature and has nothing to do with the physical temperature.

Assume $H=H_0+H_1$, where $H_0$ is diagonal in the orthonormal basis $|\underline{\sigma} \rangle$; $H_0$ and $H_1$ are Hermitian operators with real eigenvalues and orthonormal eigenvectors.
The trial wave function is represented in this representation as $|\psi(\underline{P})\rangle=\sum_{\underline{\sigma}} a(\underline{\sigma};\underline{P}) |\underline{\sigma} \rangle$.
The coefficients $a(\underline{\sigma};\underline{P})$ are complex numbers, and $|a(\underline{\sigma};\underline{P})|^2=a(\underline{\sigma};\underline{P})a^*(\underline{\sigma};\underline{P})$
is a normalized probability distribution over $\underline{\sigma}$. The average energy can be written as
\begin{eqnarray}
\langle \psi(\underline{P}) | H |\psi(\underline{P})\rangle= \sum_{\underline{\sigma}} |a(\underline{\sigma};\underline{P})|^2 [E_0(\underline{\sigma})+E_1(\underline{\sigma})],
\end{eqnarray}
where 
\begin{eqnarray}
E_0(\underline{\sigma})\equiv \langle \underline{\sigma} | H_0 |\underline{\sigma} \rangle, \hskip1cm E_1(\underline{\sigma})\equiv \mathrm{Re} \left\{ \sum_{\underline{\sigma}'} \frac{a^*(\underline{\sigma}';\underline{P})}{a^*(\underline{\sigma};\underline{P})} \langle \underline{\sigma}' | H_1 |\underline{\sigma} \rangle \right\}.
\end{eqnarray}

We are going to consider $\mu(\underline{\sigma};\underline{P})\equiv |a(\underline{\sigma};\underline{P})|^2$ as a probability measure over variables $\underline{\sigma}$ in a classical system and compute the average energies within the Bethe approximation, where the classical measure is treated as if the classical interaction graph $\mathcal{E}_c$ is a tree. For example, in the Ising model the measure is approximated by  $\mu_{BP}=\prod_i\mu_i(\sigma_i)\prod_{(ij)\in \mathcal{E}_c} \left(\mu_{ij}(\sigma_i,\sigma_j)/[\mu_i(\sigma_i)\mu_j(\sigma_j)]\right)$, given the local marginals $\mu_i(\sigma_i)$ and $\mu_{ij}(\sigma_i,\sigma_j)$. The free energy in the Bethe approximation can be written in terms of the local free-energy shifts $\Delta F_i$ and $\Delta F_{ij}$, corresponding to the changes in the free energy by adding spin $i$ and interaction between spins $i$ and $j$, respectively.  To compute these free energies we need the cavity marginals $\mu_{i \to j}(\sigma_i)$, i.e., the probability of having spin $i$ in state $\sigma_i$ in absence of the interaction with spin $j$. The equations governing these local cavity marginals are called the belief propagation (BP) equations \cite{KFL-inform-2001}. Having the cavity marginals one can obtain the Bethe estimation of the local marginals for any subset of the variables. Notice that in writing the BP equations one assumes the classical system is in a replica-symmetric (RS) phase. Even in a replica-symmetry-broken (RSB) phase, the approximation is still valid in a single pure state thanks to the exponential decay of the correlations \cite{MM-book-2009}. 

The average of any local quantity like $E_0$ and $E_1$ can be written as a function of the BP cavity marginals (or messages). Therefore, the optimization problem reads
\begin{eqnarray}
\mathcal{Z}=\sum_{\underline{P}}\sum_{\mu} e^{-\beta_{opt} \langle E_0 \rangle_{\mu}-\beta_{opt} \langle E_1 \rangle_{\mu}}I_{BP},
\end{eqnarray}
where the indicator function $I_{BP}$ ensures that the messages $\mu$ satisfy the BP equations. 
In the case of multiple BP fixed points the above partition function
would be concentrated on the one of minimum average energy for $\beta_{opt} \to \infty$. More accurate 
average energies are obtained, of course, by considering replica symmetry breaking and working with a probability 
distribution of the BP fixed points.

\section{Quantum Ising model}\label{S2}
As an example in the following we consider the quantum Ising model with Hamiltonian $H=H_0+H_1$ where $H_0=-\sum_{(ij) \in \mathcal{E}_q} J_{ij} \sigma_i^z \sigma_j^z$ and
$H_1=-\sum_{i=1}^N h_{i} \sigma_i^x$. The interaction graph is defined by set $\mathcal{E}_q$, and $N$ is the size of the system. The $\sigma^{x,y,z}$ are the standard Pauli matrices. 
Here states $|\underline{\sigma} \rangle$ are the $2^N$ configurations of the $\underline{\sigma}^z$ spins. In this case $\langle \underline{\sigma} | H_0 |\underline{\sigma} \rangle=-\sum_{(ij) \in \mathcal{E}_q} J_{ij} \sigma_i \sigma_j$ and
$\langle \underline{\sigma}' | H_1 |\underline{\sigma} \rangle=-\sum_{i} h_{i} \delta_{\sigma_i,-\sigma'_i} \delta_{\underline{\sigma}\setminus i,\underline{\sigma}'\setminus i}$.
For the trial wave functions we take the Ising ansatz:
\begin{eqnarray}
a(\underline{\sigma};\underline{P})=\frac{e^{\sum_{i} B_i \sigma_i+\sum_{(ij)\in \mathcal{E}_c} K_{ij} \sigma_i \sigma_j}}{\left( \sum_{\underline{\sigma}}e^{\sum_{i} 2B_i^R \sigma_i+\sum_{(ij)\in \mathcal{E}_c} 2K_{ij}^R \sigma_i \sigma_j} \right)^{1/2}},
\end{eqnarray}
with complex parameters $\underline{P}=\{B_i, K_{ij}|i=1,\dots,N,(ij)\in \mathcal{E}_c\}$. By $B_i^R$ and $K_{ij}^R$ we mean the real part of the parameters. This results in the Gibbs measure  $\mu(\underline{\sigma};\underline{P})=|a(\underline{\sigma};\underline{P})|^2$ of a classical spin-glass model with external fields $2B_i^R$ and couplings $2K_{ij}^R$.
Notice that the classical interaction graph $\mathcal{E}_c$ could be different from the quantum one $\mathcal{E}_q$. For simplicity, in the following we will assume that the two coincide as happens in zero transverse fields; better representations could be obtained by adding the higher order neighbors to $\mathcal{E}_c$. 

\begin{figure}
\includegraphics[width=6cm,height=10cm]{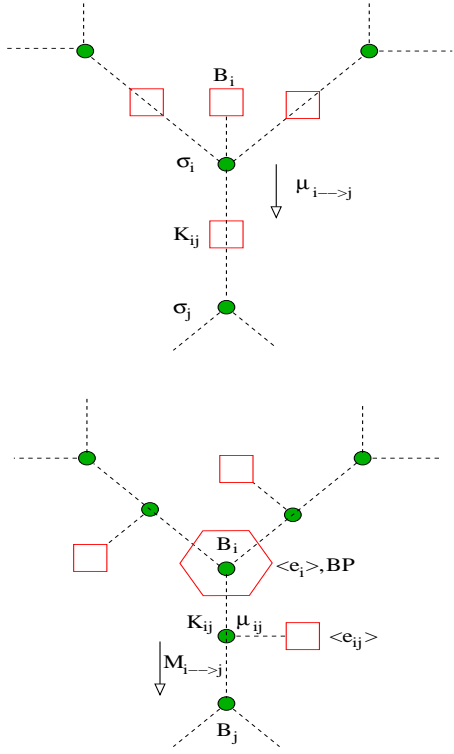}
\caption{(top) The interaction graphs of the classical system  representing the trial wave function and (bottom) the resulting variational problem. The variables are shown with solid circles, and the interactions are shown with open polygons.}\label{f1}
\end{figure}

Given the classical measure, we write the BP equations for the cavity marginal $\mu_{i \to j}(\sigma_i)$ of spin $i$ in the absence of spin $j$:
\begin{eqnarray}
\mu_{i \to j}(\sigma_i) \propto e^{2B_i^R \sigma_i}\prod_{k \in \partial i \setminus j} \left(\sum_{\sigma_k} e^{2K_{ik}^R \sigma_i\sigma_k }\mu_{k \to i}(\sigma_k) \right)\equiv \mathcal{BP}_{i\to j},
\end{eqnarray}
where $\partial i$ refers to the set of spins interacting with spin $i$ in $\mathcal{E}_c$. These equations can easily be obtained by assuming a tree structure for the classical interaction graph \cite{MM-book-2009}. Figure \ref{f1} displays the set of variables and interactions in the classical interaction graph. Having the cavity marginals, the average of local energies $e_{ij} \equiv -J_{ij} \sigma_i^z\sigma_j^z$  and  $e_{i}\equiv  -h_i \sigma_i^x $ read
\begin{align}
\langle e_{ij} \rangle_{\mu} &=-J_{ij} \sum_{\sigma_i,\sigma_j} \sigma_i\sigma_j \mu_{ij}(\sigma_i,\sigma_j), \label{eij}\\ 
\langle e_{i} \rangle_{\mu} &= -h_i \sum_{\sigma_i,\sigma_{\partial i}}  e^{-2B_i^R\sigma_i-\sum_{j\in \partial i} 2K_{ij}^R\sigma_i\sigma_j } \cos\left(2B_i^I\sigma_i+\sum_{j\in \partial i}2K_{ij}^I\sigma_i\sigma_j \right)\mu_{i,\partial i}(\sigma_i,\sigma_{\partial i}) \label{ei},
\end{align}
with the local BP marginals $\mu_{ij}(\sigma_i,\sigma_j)\propto e^{2K_{ij}^R \sigma_i\sigma_j}\mu_{i \to j}(\sigma_i)\mu_{j \to i}(\sigma_j)$ and $\mu_{i,\partial i}(\sigma_i,\sigma_{\partial i})\propto e^{2B_{i}^R \sigma_i}\prod_{j \in \partial i}[e^{2K_{ij}^R \sigma_i\sigma_j}\mu_{j \to i}(\sigma_j)]$. We see that the only dependence of the total average energy on the imaginary part of the parameters comes explicitly from the $\langle e_{i} \rangle_{\mu}$. Consequently, for $h_i \ge 0$ and $h_i <0$ we can minimize the total average energy by setting $B_i^I=0, K_{ij}^I=0$ and $B_i^I=\pi/2, K_{ij}^I=0$, respectively. Without loss of generality, in the following we assume the $h_i\ge 0$ and the imaginary parameters are zero.

The above average energies define the Boltzmann weight $e^{-\beta_{opt} \langle E_0 \rangle_{\mu}-\beta_{opt} \langle E_1 \rangle_{\mu}}$ for a given configuration of the variational parameters and the BP messages; see figure \ref{f1}.  
The cavity marginals of the parameters (including the BP messages) can be written in a higher-level 
Bethe approximation, resembling the one-step RSB equations \cite{MM-book-2009}: 
\begin{eqnarray}
M_{i \to j}(K_{ij},\mu_{ij}) \propto  e^{-\beta_{opt} \langle e_{ij} \rangle_{\mu}}
\sum_{B_i,\{K_{ik},\mu_{ik}| k \in \partial i \setminus j\}} e^{-\beta_{opt} \langle e_i \rangle_{\mu} } \prod_{k \in \partial i \setminus j} M_{k \to i}(K_{ik},\mu_{ik}) I_{BP}(i),
\end{eqnarray}
where for brevity we defined $\mu_{ij}\equiv \{\mu_{i\to j},\mu_{j\to i}\}$ and $I_{BP}(i)\equiv \prod_{j \in \partial i} \delta(\mu_{i\to j}-\mathcal{BP}_{i\to j})$.
The Bethe free energy is given by $F= \sum_i \Delta F_i-\sum_{(ij) \in \mathcal{E}_c} \Delta F_{ij}$ with
\begin{align}
e^{-\beta_{opt} \Delta F_i} &\equiv \sum_{B_i,\{K_{ij},\mu_{ij}|j\in \partial i\}} e^{ -\beta_{opt} \langle e_i \rangle_{\mu} } \prod_{j \in \partial i} M_{j \to i}(K_{ij},\mu_{ij}) I_{BP}(i), \\
e^{-\beta_{opt} \Delta F_{ij}} &\equiv \sum_{K_{ij},\mu_{ij}} e^{+\beta_{opt} \langle e_{ij} \rangle_{\mu} } M_{i \to j}(K_{ij},\mu_{ij})M_{j \to i}(K_{ij},\mu_{ij}),  
\end{align}
where $\Delta F_i$ and $\Delta F_{ij}$ are the free-energy shifts by adding node $i$ and link $(ij)$, respectively. Notice that in the last equation we have the positive sign in the exponential to count correctly the energy contribution $\langle e_{ij} \rangle_{\mu}$.

The $\beta_{opt} \to \infty$ limit of the above equations, 
taking the scaling $M_{i \to j}(K_{ij},\mu_{ij})=e^{\beta_{opt} \mathcal{M}_{i \to j}(K_{ij},\mu_{ij})}$, read 
\begin{align}
\mathcal{M}_{i \to j}(K_{ij},\mu_{ij}) &= -\langle e_{ij} \rangle_{\mu}+ 
\max_{B_i,\{K_{ik},\mu_{ik}| k \in \partial i \setminus j\}:I_{BP}(i)} 
\left\{ -\langle e_i \rangle_{\mu} + \sum_{k \in \partial i \setminus j} \mathcal{M}_{k \to i}(K_{ik},\mu_{ik}) \right\},
\end{align} 
which in short we call the MaxSum-BP equations \cite{ABRZ-prl-2011,ABRZ-jstat-2011}. The equations can be solved by iteration starting from random initial messages. In each iteration we have to shift $\mathcal{M}_{i \to j}$ by a constant to keep $\max_{K_{ij},\mu_{ij}}\mathcal{M}_{i \to j}(K_{ij},\mu_{ij})=0$.   
Finally, the minimum energy is given by $E_g=\lim_{\beta_{opt} \to \infty} F = - (\sum_i \Delta e_i-\sum_{(ij) \in \mathcal{E}_c} \Delta e_{ij})$
with the local energy shifts
\begin{align}
\Delta e_i &\equiv \max_{B_i,\{K_{ij},\mu_{ij}|j\in \partial i\}:I_{BP}(i)} \left\{ -\langle e_i \rangle_{\mu} + \sum_{j \in \partial i} \mathcal{M}_{j \to i}(K_{ij},\mu_{ij}) \right\}, \\
\Delta e_{ij} &\equiv \max_{K_{ij},\mu_{ij}} \left\{ \langle e_{ij} \rangle_{\mu} + \mathcal{M}_{i \to j}(K_{ij},\mu_{ij})+  \mathcal{M}_{j \to i}(K_{ij},\mu_{ij}) \right\}.  
\end{align}
Before solving the above equations we shall consider some simpler cases.

\subsection{Zero couplings: Mean field solution}\label{S21} 
In the zeroth order of the approximation we take $K_{ij}=0$ for any $(ij)$. This is a mean-field approximation with a factorized measure $\mu(\underline{\sigma})=\prod_i \mu_i(\sigma_i)$, where $\mu_i(\sigma_i)=e^{2B_i\sigma_i}/[2\cosh(2B_i)]$. Then using equations \ref{eij} and \ref{ei} we obtain the average local energies: $\langle e_{ij}\rangle_{\mu}=-J_{ij} \tanh(2B_i)\tanh(2B_j)$ and $\langle e_{i}\rangle_{\mu}=-h_i/\cosh(2B_i)$; therefore
\begin{align}
\langle \psi(\underline{B}) | H |\psi(\underline{B})\rangle= -\sum_{(ij) \in \mathcal{E}_q} J_{ij} \tanh(2B_i)\tanh(2B_j)
-\sum_{i} \frac{h_i}{\cosh(2B_i)}. 
\end{align}
Here we write directly the MaxSum equations that can be used to estimate the optimal parameters and the minimum average energy:
\begin{align}
\mathcal{M}_{i \to j}(B_i) &= -\langle e_{i} \rangle_{\mu}+ \sum_{k\in \partial i\setminus j}
\max_{ B_k } 
\left\{ -\langle e_{ik} \rangle_{\mu} + \mathcal{M}_{k \to i}(B_k) \right\}.
\end{align} 
Then we find the optimal paramters by maximizing the local MaxSum weights: 
\begin{align}
B_i^*=\arg \max_{B_i} \left\{-\langle e_{i} \rangle_{\mu}+ \sum_{j\in \partial i}
\max_{ B_j } 
\left\{ -\langle e_{ij} \rangle_{\mu} + \mathcal{M}_{j \to i}(B_j) \right\} \right\}.
\end{align}

\subsection{Zero fields: Symmetric solution}\label{S22}
As long as the fields $B_i$ are zero we have always a symmetric solution $\mu_{i \to j}(\sigma_i)=1/2$ to the BP equations in the classical system. This gives the average local energies: $\langle e_{ij}\rangle_{\mu}=-J_{ij} \tanh(2K_{ij})$ and $\langle e_{i}\rangle_{\mu}=-h_i/[\prod_{j\in \partial i} \cosh(2K_{ij})]$, and 
\begin{align}
\langle \psi(\underline{K}) | H |\psi(\underline{K})\rangle= -\sum_{(ij) \in \mathcal{E}_q} J_{ij} \tanh(2K_{ij})
-\sum_{i} \frac{h_i}{\prod_{j\in \partial i} \cosh(2K_{ij})}. 
\end{align}
The resulting MaxSum equations are
\begin{align}
\mathcal{M}_{i \to j}(K_{ij}) &= -\langle e_{ij} \rangle_{\mu}+ 
\max_{\{K_{ik}| k \in \partial i \setminus j\}} 
\left\{ -\langle e_i \rangle_{\mu} + \sum_{k \in \partial i \setminus j} \mathcal{M}_{k \to i}(K_{ik}) \right\},
\end{align} 
and the optimal couplings are estimated by 
\begin{eqnarray}
K_{ij}^*= \arg \max_{K_{ij}}\left\{ \langle e_{ij} \rangle_{\mu} + \mathcal{M}_{i \to j}(K_{ij})+  \mathcal{M}_{j \to i}(K_{ij})\right\}.  
\end{eqnarray}

Notice that here we have $\langle \sigma_i^z \rangle=0$, which is not the case in the ordered phase. 
In addition, the symmetric solution does not give an accurate average energy when replica symmetry is broken, which may happen
for large couplings in the classical system; the Bethe approximation works well when distant spins are nearly independent, whereas 
the symmetric solution does not respect this property in an RSB phase. To get around this problem one can work with the nontrivial 
BP fixed points, e.g., by demanding a total magnetization of magnitude greater than $\delta m \ll 1$. 
At the same time one may need to limit the range of couplings to $|K_{ij}|<K_{max}$ in order to avoid dominance by very large couplings.

\subsection{General solution}\label{S23}
In general to solve the MaxSum-BP equations we have to work with discrete fields $B_l\in \{l\delta B|l=-L_B,\dots,L_B\}$, 
couplings $K_l\in \{l\delta K|l=-L_K,\dots,L_K\}$, and BP cavity fields $\nu_l\in \{l\delta \nu|l=-L_{\nu},\dots,L_{\nu}\}$. 
The BP cavity fields $\nu_{i \to j}$ are defined by $\mu_{i \to j}(\sigma_i)\propto e^{\nu_{i \to j}\sigma_i}$.
An exhaustive solution of the MaxSum-BP equations would take a time of order $Nd(2L_B)[(2L_K) (2L_{\nu})]^{d}$ where $d$ is the maximum degree
in $\mathcal{E}_c$. Notice that given the couplings $K_{ij}$ and the input BP messages $\nu_{j \to i}$ around spin $i$,
one obtains $\nu_{i \to j}$ for any value of $B_i^R$ from the BP equations.
The above equations can be solved more efficiently (for large degrees) by using a convolution function of four variables 
(needed to compute $\langle e_i \rangle_{\mu}$ in the MaxSum-BP equations) resulting in a time complexity of order
$Nd(2L_K)(2L_{\nu})^6$; see the Appendix. In the following instead we use a computationally easier but approximate way of solving 
the equations by restricting the domain of variables:   
We start by assigning a small number of randomly selected states $S_{ij}=\{(K_{ij}^1,\nu_{ij}^1),\dots,(K_{ij}^S,\nu_{ij}^S)\}$ 
to each variable $(K_{ij},\nu_{ij})$.
Then we run the MaxSum-BP equations with these restricted search spaces to converge the equations and sort the states in $S_{ij}$ 
according to their MaxSum-BP weights:
\begin{eqnarray}
w_{ij}(K_{ij}^l,\nu_{ij}^l)=  \langle e_{ij} \rangle_{\mu} + \mathcal{M}_{i \to j}(K_{ij}^l,\nu_{ij}^l)+  \mathcal{M}_{j \to i}(K_{ij}^l,\nu_{ij}^l).  
\end{eqnarray}
Next we update the search spaces by replacing the states having smaller weights with some other ones generated randomly but 
close to the best state observed during the algorithm. The above two steps are repeated to find better search spaces and therefore parameters. 
The algorithm performance would depend on the size of the search spaces, approaching the correct one for $S \to \infty$. 

The search spaces at the beginning are chosen randomly therefore we introduce a tolerance $\delta \nu$ to accept those BP messages 
that satisfy the BP equations within $\pm \delta \nu$. One may start from a large tolerance and decrease it slowly after each 
update of the search spaces. The mean-field solution mentioned before can provide a good initial point for the search spaces.

\subsection{Numerical results}\label{S24}
\begin{figure}
\includegraphics[width=10cm]{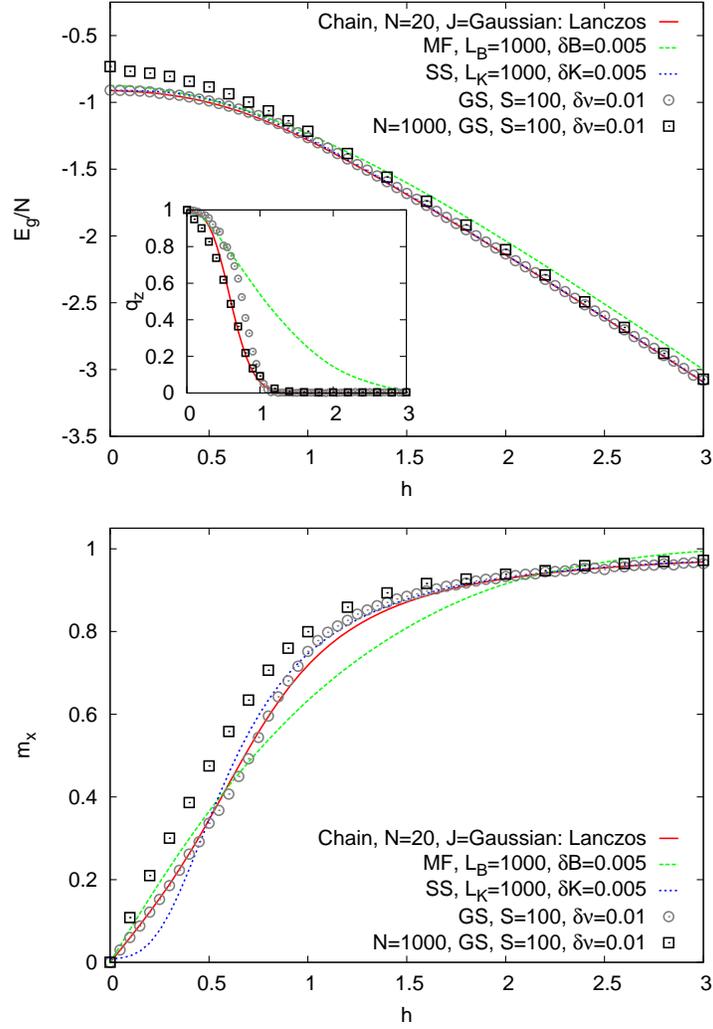}
\caption{Comparing the exact results with those of the mean-field (MF), the symmetric solution (SS) and the general solution (GS) 
in a single instance of the quantum Ising model in transverse field $h$ on a chain of size $N=20$. For comparison we also 
display the GS results for $N=1000$. Here $E_g$ and $m_x$ are the ground-state energy and  magnetization in the $x$ direction, 
respectively. The inset shows the Edwards-Anderson order parameter $q_z=(\sum_i \langle \sigma_i^z \rangle ^2)/N$. 
The couplings $J_{ij}$ are Gaussian random numbers of mean zero and variance one. $\delta B$, $\delta K$, and $\delta \nu$ are 
the sizes of bins in the discrete representation of the parameters, and $S$ is the number of states in the restricted domains. The data 
for GS are obtained by restricting the search algorithm to total magnetizations of magnitude greater than $\delta m=0.05$.}\label{f2}
\end{figure}

First, we consider a spin chain in uniform and positive transverse field but with random couplings.
In figure \ref{f2} we compare the results obtained by the above approximations with the exact ones obtained by the modified 
Lanczos method \cite{GDMA-prb-1986}.  As expected, the mean-field ansatz is better than the symmetric solution in the ordered 
phase where the local magnetizations are nonzero. The reverse happens in the disordered phase and even in the ordered phase close 
to the transition point where the local magnetizations are still small. The two limiting behaviors are therefore displayed in the
general solution.

\begin{figure}
\includegraphics[width=10cm]{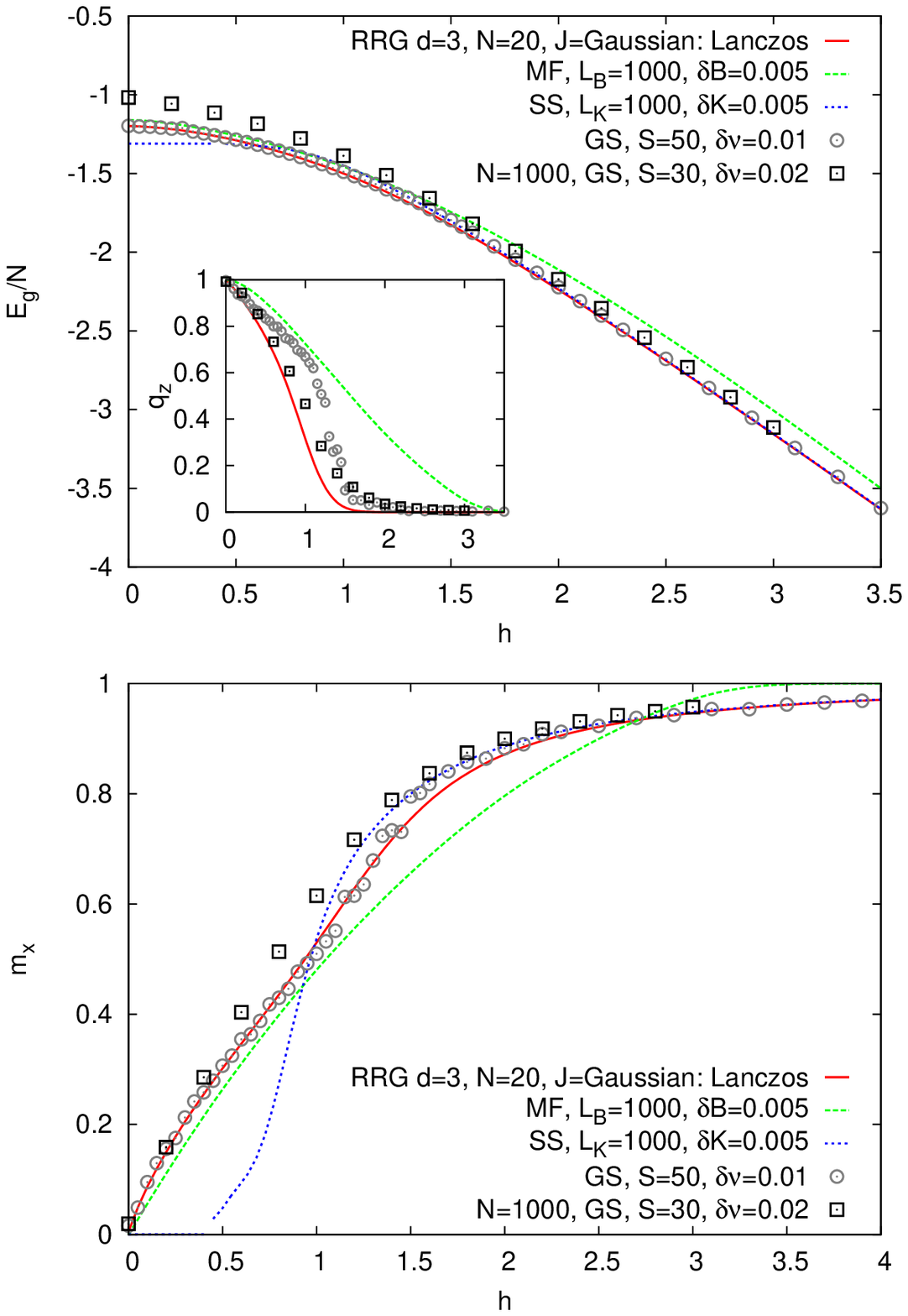}
\caption{The quantum Ising model in transverse field $h$ on a random regular graph (RRG) of degree $d=3$ and size $N=20$. 
For comparison we also display the GS results for $N=1000$. The couplings $J_{ij}$ are Gaussian random numbers of mean zero 
and variance one.
The data for GS are obtained by restricting the search algorithm to total magnetizations of magnitude greater than
$\delta m=0.05$ and couplings of magnitude less than $K_{max}=1$.}\label{f3}
\end{figure}

As another example we study the same model on a single instance of random regular graphs where each spin interacts with a fixed 
number of other randomly selected spins. The results displayed in figure \ref{f3} show similar behaviors observed above, 
except that in the ordered phase the symmetric solution gives a lower ground state energy than the exact one. As explained before, 
this is due to the poor estimation of the average energy by the symmetric solution when there are multiple BP fixed points. 
Moreover, due to the loops and small size of the system we find larger deviations from the exact data compared to the chain model.

\begin{figure}
\includegraphics[width=10cm]{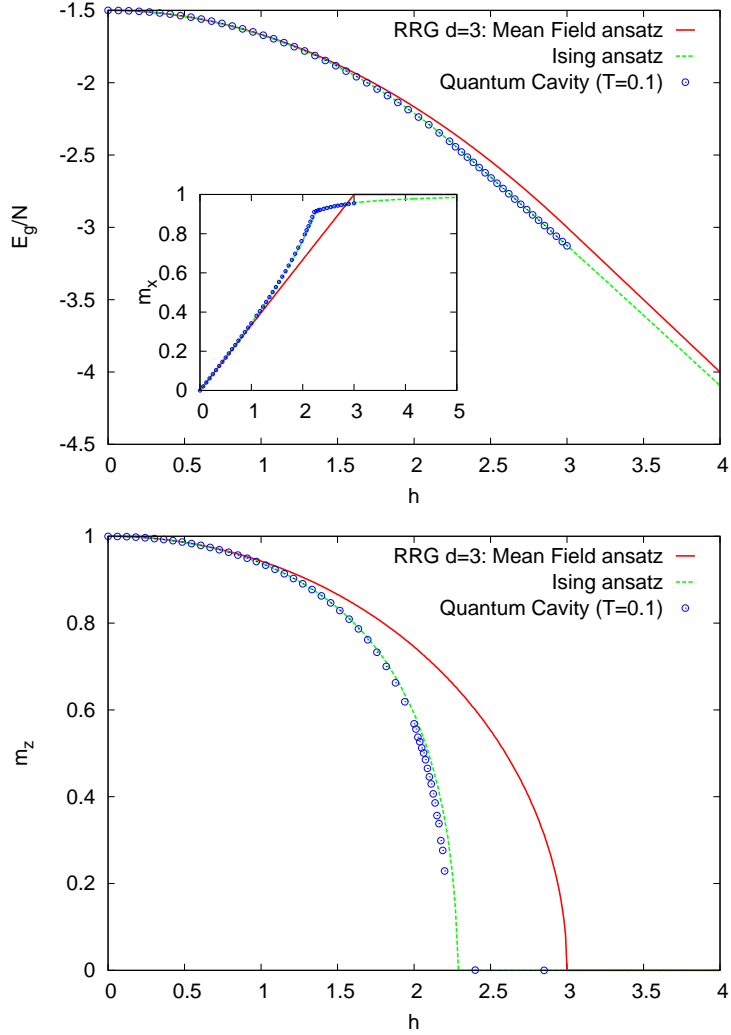}
\caption{The quantum Ising model with ferromagnetic interactions ($J_{ij}=1$) in transverse field $h$ on RRG of degree $d=3$. 
The results of homogeneous mean field $B_i=B$ and Ising ($B_i=B, K_{ij}=K$) trial wave functions are compared with the results of the quantum cavity method \cite{KRSZ-prb-2008} for a small temperature in the thermodynamic limit.}\label{f4}
\end{figure}

Similar qualitative behaviors are observed in ferromagnetic ($J_{ij}=1$) and $\pm J$ spin-glass ($J_{ij}=\pm 1$ with equal
probability) models on a random regular graph of degree $d=3$. In the ferromagnetic case, a reasonable trial wave function is obtained by taking $B_i=B$ and $K_{ij}=K$. This simplification allows us to find easily the optimal parameters, and so the critical field $h_c^{ferro}\sim 2.29$ in the thermodynamic limit. Figure \ref{f4} displays the phase diagram of the ferromagnetic model obtained in this way. 
For the $\pm J$ spin-glass model, the general solution on single instances of size $N=1000$ gives $h_c^{\pm J} \simeq 2.0$.
The corresponding values in the thermodynamic limit given in Refs. \cite{KRSZ-prb-2008} and \cite{LSS-prb-2008} are:
$h_c^{ferro} \sim 2.23$ and $h_c^{\pm J} \sim 1.77$.

\section{Conclusion}\label{S3}
We suggested a heuristic message-passing algorithm to find an approximated ground state of a quantum system for 
a given instance of (possibly disordered) couplings. 
This was done by exploiting an efficient representation of the wave function and relating the quantum state of 
the system to the Gibbs state in a classical system.
The local and distributive nature of the algorithm could help us to study the ground-state properties of large-scale 
quantum problems in a parallel computation. 
We used the following main approximations to make the study simple and clear: (i) working at most with the two-body 
interactions in the classical system, (ii) assuming the same classical and quantum interaction graphs, and  
(iii) estimating the classical average energies within the replica-symmetric approximation. 
Our approach, however, allows for a systematic way of improving the algorithm by considering better trial wave functions,  using
more accurate cluster variational or generalized BP approximations \cite{K-pr-1951,YFW-nips-2001}, 
and taking the effects of replica symmetry breaking into account. 
Finally, the method can be applied to other interesting quantum systems; in a forthcoming paper we will show how 
this works in the Hubbard model in the presence of the sign problem.

\acknowledgments
I would like to thank S. Lal, G. Semerjian, and R. Zecchina for reading the manuscript and their constructive comments and A. Montorsi, and M. Roncaglia for helpful discussions. I thank G. Semerjian also for providing us with the quantum cavity data. Support from ERC  Grant No. OPTINF  267915 is acknowledged.

\appendix

\section{Solving the MaxSum-BP equations}\label{app}
Let us write the BP equations in terms of the cavity fields:
\begin{eqnarray}
\nu_{i \to j}=2B_i+\frac{1}{2}\sum_{k \in \partial i \setminus j} \ln[ \frac{\cosh(\nu_{k \to i}+2K_{ik})}{\cosh(\nu_{k \to i}-2K_{ik})}].
\end{eqnarray}
Using Eqs. \ref{eij} and \ref{ei} we rewrite the local average energies 
\begin{align}\label{ei-eij-app}
\langle e_{ij} \rangle_{\mu} &=-J_{ij} \frac{e^{2K_{ij}}\cosh(\nu_{i\to j}+\nu_{j\to i})-e^{-2K_{ij}}\cosh(\nu_{i\to j}-\nu_{j\to i})}{e^{2K_{ij}}\cosh(\nu_{i\to j}+\nu_{j\to i})+e^{-2K_{ij}}\cosh(\nu_{i\to j}-\nu_{j\to i})}, \\ 
\langle e_{i} \rangle_{\mu} &= - \frac{2h_i}{e^{2B_i}y_{i,+}+e^{-2B_i}y_{i,-}},
\end{align}
where
\begin{align}
y_{i,+} &=\prod_{k\in \partial i } \frac{\cosh(\nu_{k \to i}+2K_{ik})}{\cosh(\nu_{k \to i})}, \\ 
y_{i,-} &=\prod_{k\in \partial i } \frac{\cosh(\nu_{k \to i}-2K_{ik})}{\cosh(\nu_{k \to i})}.
\end{align} 
So the MaxSum-BP equations read 
\begin{multline}
\mathcal{M}_{i \to j}(K_{ij},\nu_{ij}) = J_{ij}\frac{e^{2K_{ij}}\cosh(\nu_{i\to j}+\nu_{j\to i})-e^{-2K_{ij}}\cosh(\nu_{i\to j}-\nu_{j\to i})}{e^{2K_{ij}}\cosh(\nu_{i\to j}+\nu_{j\to i})+e^{-2K_{ij}}\cosh(\nu_{i\to j}-\nu_{j\to i})} \\ 
+\max_{B_i,\{K_{ik},\nu_{ik}| k \in \partial i \setminus j\}: I_{BP}(i)} \Bigg\{  \frac{2h_i}{e^{2B_i}y_{i,+}+e^{-2B_i}y_{i,-}} + \sum_{k \in \partial i \setminus j} \mathcal{M}_{k \to i}(K_{ik},\nu_{ik}) \Bigg\}.
\end{multline}
To split the maximum over the set $\{K_{ik},\nu_{ik}| k \in \partial i \setminus j\}$ we introduce 
a convolution function $F_{t}(x_+^{t},x_-^{t},y_+^{t},y_-^{t})$ to keep track of the quantities needed to compute the average local energies
and satisfy the BP equations. At each step $t$ we take the maximum over the subset $\{K_{ik_t},\nu_{ik_t}\}$ for $t=1,\dots,d_i-1$, where $d_i$ is the 
degree of node $i$ in $\mathcal{E}_c$, and the variables are  
\begin{align}
x_{+}^t &=\frac{1}{2}\sum_{t'\le t }  \ln[ \frac{\cosh(\nu_{k_{t'} \to i}+2K_{ik_{t'}})}{\cosh(\nu_{k_{t'} \to i}-2K_{ik_{t'}})} ], \\ 
x_{-}^t &=\frac{1}{2}\sum_{t'> t }  \ln[ \frac{\cosh(\nu_{k_{t'} \to i}+2K_{ik_{t'}})}{\cosh(\nu_{k_{t'} \to i}-2K_{ik_{t'}})} ], \\ 
y_{+}^t &=\prod_{t'\le t } \frac{\cosh(\nu_{k_{t'} \to i}+2K_{ik_{t'}})}{\cosh(\nu_{k_{t'} \to i})}, \\ 
y_{-}^t &=\prod_{t'\le t } \frac{\cosh(\nu_{k_{t'} \to i}-2K_{ik_{t'}})}{\cosh(\nu_{k_{t'} \to i})}.
\end{align}
The convolution function is updated sequentially as
\begin{multline}
F_{t+1}(x_+^{t+1},x_-^{t+1},y_+^{t+1},y_-^{t+1})= \\
\max_{K_{ik_{t+1}}, \nu_{ik_{t+1}}:I_{t+1}} \Bigg\{ F_{t}(x_+^t,x_-^t,y_+^t,y_-^t)+\mathcal{M}_{k_{t+1} \to i}(K_{ik_{t+1}},\nu_{ik_{t+1}}) \Bigg\},
\end{multline}
with the set of constraints $I_{t+1}$:
\begin{align}
x_+^{t+1} &=x_+^t+ \frac{1}{2}\ln[\frac{\cosh(\nu_{k_{t+1} \to i}+2K_{ik_{t+1}})}{\cosh(\nu_{k_{t+1} \to i}-2K_{ik_{t+1}})}], \\ 
x_-^{t+1} &=x_-^t- \frac{1}{2}\ln[\frac{\cosh(\nu_{k_{t+1} \to i}+2K_{ik_{t+1}})}{\cosh(\nu_{k_{t+1} \to i}-2K_{ik_{t+1}})}] = \nu_{i \to k_{t+1}}-2B_i-x_+^{t}, \\ 
y_+^{t+1} &=y_+^t  \frac{\cosh(\nu_{k_{t+1} \to i}+2K_{ik_{t+1}})}{\cosh(\nu_{k_{t+1} \to i})}, \\ 
y_-^{t+1} &=y_-^t  \frac{\cosh(\nu_{k_{t+1} \to i}-2K_{ik_{t+1}})}{\cosh(\nu_{k_{t+1} \to i})}, 
\end{align}
and the following boundary condition: $F_{0}(0,x_-,1,1)=0$; otherwise, it is $-\infty$. Finally, the MaxSum-BP message is computed after $t=d_i-1$ steps by
\begin{align}
\mathcal{M}_{i \to j}(K_{ij},\nu_{ij}) = J_{ij}\frac{e^{2K_{ij}}\cosh(\nu_{i\to j}+\nu_{j\to i})-e^{-2K_{ij}}\cosh(\nu_{i\to j}-\nu_{j\to i})}{e^{2K_{ij}}\cosh(\nu_{i\to j}+\nu_{j\to i})+e^{-2K_{ij}}\cosh(\nu_{i\to j}-\nu_{j\to i})} \\ 
+\max_{B_i, y_+^t, y_-^t} \Bigg\{ \frac{2 h_i}{e^{2B_i}y_+^t+e^{-2B_i}y_-^t} + F_{t}(\nu_{i \to j}-2B_i,\nu_{j \to i},y_+^t,y_-^t) \Bigg\},
\end{align}

\end{document}